\documentclass[aps,amssymb,amsmath,superscriptaddress,twocolumn,altaffilletter]{revtex4}

\usepackage{amsmath,amsfonts,amsthm,amssymb,amsxtra,dsfont}
\usepackage{graphicx}
\usepackage{amsfonts}
\usepackage{amsmath,epsf,epic}
\usepackage{tikz}%\usetikzlibrary{arrows,shapes,trees,..}
\usepackage{float}
 \usepackage[title]{appendix} 

\theoremstyle{definition}

\theoremstyle{remark}

\renewcommand{\phi}{\varphi}

\newcommand{\eps}{\epsilon}

\DeclareMathOperator{\sgn}{sgn}

\renewcommand{\vec}[1]{\mathbf{#1}}

\newcommand{\be}{\begin{equation}}
\newcommand{\ee}{\end{equation}}
\newcommand{\bea}{\begin{eqnarray}}
\newcommand{\eea}{\end{eqnarray}}

\def\H{H}

\def\Heff{\H_{\text{eff}}}

\def\Hel{\H_{\text{el}}}

\newcommand{\Ii}{\mathrm i} % complex unit 
\newcommand{\Ee}{\mathrm e} % Euler's number

\newcommand{\DInt}[2][\,]{{\mathrm d}^{\hspace{-0.20ex}#1}\hspace{-0.25ex}#2}

\newcommand{\abs}[1]{\left |#1 \right|}

\def\RL{\text{RL}}
\def\El#1{{\rm #1}} % Elements in Chemical Formulas
\def\SpDot{\cdot} % notation for scalar product
\def\FBZ{\text{FBZ}}

\usepackage{lipsum}
\newcommand{\omegalo}{\omega_0}
\newcommand{\dd}{d}
\begin{document}

\author{Maximilian Duell}
\email{duell@math.lmu.de}
 \author{Christian Hainzl} 
 \email{hainzl@math.lmu.de}
 \affiliation{Mathematisches Institut, Ludwig-Maximilians-Universit\"at M\"unchen, 80333 Munich, Germany}
 \author{Eman Hamza}
 \email{hamza@math.lmu.de}
 \affiliation{Mathematisches Institut, Ludwig-Maximilians-Universit\"at M\"unchen, 80333 Munich, Germany}
 \affiliation{Faculty of Science, Cairo University, Cairo 12613, Egypt}

\title{Enhanced pairing mechanism in Cuprate-type crystals}

\begin{abstract}
Using a BCS mean-field approach, we show how the interplay between low-momentum optical phonons and Jahn-Teller-type lattice distortions can open an attractive channel that allows the formation of pairs with the corresponding density exhibiting characteristic features of a pair-density wave (PDW). We demonstrate this numerically on a copper-oxide type lattice. 
\end{abstract}

\maketitle

While the pairing mechanism in conventional superconductors has long been well
understood, the situation for cuprate superconductors is still controversial and
unexplained thirty-five years after their discovery.  Although the traditional
phonon-mediated BCS pairing mechanism has been largely ruled out as the main cause of
high-temperature superconductivity,  several experimental groups, e.g.~\cite{Lanzara, G}, reported observations of sufficiently strong interactions
between certain optical modes and doped charge carriers. A number of recent
experiments \cite{He, Lee} further suggest a pronounced correlation between the
superconducting gap and the strength of electron-phonon coupling at small
momentum transfer \cite{Y, DCSN}. Bednorz and M\"uller \cite{BM} were motivated
in their search for new superconducting materials by the idea that lattice
distortions in the sense of dynamic Jahn-Teller polarons could be the novel glue
for electron pairing, much stronger than the conventional BCS pairing mechanism
\cite{M1,HN}. In light of their sensational success, it seems perfectly
reasonable to assume that this fundamental discovery of copper oxide
superconductors was no coincidence, but rather confirmation of the fact that
strong dynamic lattice distortions are required to achieve high values of $T_c$.
Such dynamic distortions undoubtedly seem to play a role in cuprates
\cite{KBM,KB}. The aim of the present work is to present a previously unconsidered  pairing mechanism
driven by a synergy of Jahn-Teller type crystal lattice deformations and low-momentum optical
phonon vibrations.

In a recent paper, one of the authors (C.H.) and M. Loss \cite{HL} pointed out
that for interactions more general than depending only on relative distance,
arbitrary electron pairs with momenta $(\vec k, \vec k')$ and equal energy
$\eps(\vec k)=\eps(\vec k')$ can lead to instability of the Fermi sea. With this
in mind, one is lead to consider pairs $(\vec k, \vec k')$ such that
$\frac{|\vec k-\vec k'|}{|\vec k_{\rm F}|} \ll 1$ with both momenta close to the
Fermi surface.

%For reasons of simplicity we shall restrict ourselves here 
We will show that it is further sufficient to consider
pairs with equal momentum and
opposite spin, and in this case a remarkably simple and explicitly
solvable model is obtained. Therein the pair-forming effective interactions
result from the above mentioned combination of optical phonon interactions and
lattice deformations.
Interestingly, with this restriction to pairs of the form $(\vec k,\vec k)$, the
corresponding gap equation has a simple structure.  Most notably, the critical
temperature depends linearly on the interaction strength. We will describe
numerical results justifying this restriction using an example potential with non-vanishing
momentum transfer.

Let us now become more concrete. We consider a diatomic copper oxide lattice
(see Figure \ref{fig:lattice}). Using the Wegner flow method \cite{W1,W2}, we
obtain  an effective interaction between charge carriers, similar to the earlier
derivations of Fr\"ohlich \cite{F1}, and Bardeen-Pines \cite{BP}. The exact form
of the effective interaction depends on the details of the associated Bloch
functions and hence on the details of the lattice geometry. 

Consequently, we apply the BCS approximation to the resulting Hamiltonian and
investigate the possibility of correlated pairs due to the instability of the
Fermi sea.  In other words, we consider the non-interacting Fermi gas as the
parent compound for the superconducting behavior, with the chemical potential
$\eps(\vec k_{\rm F})$ playing the role of the doping parameter.  Once we obtain
an effective interaction, we consider the resulting BCS gap equation for pairs
of the form $(\vec{k},\vec{k})$, which now takes the following simplified form
\begin{equation}\label{eq:introgap}
  \left( \frac{\sqrt{(\eps(\vec{k}) - \eps(\vec{k}_{\rm F}))^2 +
  |\Delta(\vec{k})|^2}}{\tanh\left(\frac {\sqrt{(\eps(\vec{k})
  -\eps(\vec{k}_{\rm F}))^2 + |\Delta(\vec{k})|^2})}{2T}\right)} + \frac{V(\vec{k})}{2}\right)\Delta(\vec{k}) =  0,
\end{equation}
where $\vec k$ is the crystal momentum, $\vec k_{\rm F}$ the Fermi-momentum and
$V$ is the effective interaction, with attractive component $V\leq 0$. On the one hand, our simplifications lead to the nice
equation \eqref{eq:introgap}, but on the other hand they have unfortunately
removed the phase dependence, since the solutions of \eqref{eq:introgap} are
uniquely determined only up to an arbitrary phase. For this reason,  our numerical solutions of the gap equation
is only concerned with  the absolute values of $\Delta$.   
Further, it is important to emphasize the following; if the crystal lattice is
perfectly symmetrical, then the effective interaction $V(\vec k)$ vanishes
identically. However, Jahn-Teller-type lattice distortions, which 
form dynamically in presence of charge carriers, allow non-vanishing interactions $V(\vec k)$,
which in turn open attractive channels for Cooper pairing. We present an example of such
an interaction in Section \ref{sec:model1}.  The solution~$\Delta(\vec k)$
in~\eqref{eq:introgap} is automatically concentrated near the Fermi surface, as seen
in Figure~\ref{fig:gaptight}.
 The corresponding critical pairing temperature $T^*$ has the simple form
  \begin{equation}\label{eq:introT*}
    T^*=-\frac{V(\vec{k}_{\rm F})}{4}.
  \end{equation}
Let us emphasize that the magnitude of the interaction $V(\vec{k}_{\rm F})$
depends significantly on the strength of the coupling of charge carriers to the
lattice, which according to~\eqref{eq:introT*} determines the
temperature~\(T^*\), below which the BCS approach predicts the occurrence of
correlated pairs. This is in line with the original insightful heuristics used by Bednorz
and M\"uller in their successful searches for superconducting materials. 
The linear dependence~\eqref{eq:introT*} arises as a consequence of the
simplicity of the effective gap equation governing the formation of \((\vec k, \vec
k)\) pairs and provides a strong contrast to 
the standard BCS critical temperature which is exponentially small in the coupling constant.
The distinct
behaviors of the two types of pairings can be understood by noting that the underlying approximations responsible for the linear behavior~\eqref{eq:introT*} can be justified only for~\((\vec k, \vec k)\),
while they certainly fail for \((\vec k, -\vec k)\) (see Figures~\ref{fig:kmkwf}
  and \ref{fig:kkwfa}).

 We propose the following interpretation of our work for copper oxide materials:
Since we neglect the strong Coulomb repulsion among electrons and use the
BCS mean-field approach, our analysis cannot be directly applied to the
occurrence of superconductivity itself, but it could well describe the
pseudo-gap (PG), where $T^*$ is the corresponding critical temperature. If the
chemical potential $\eps(\vec{k}_{\rm F})$ models the amount of doping, then the
phase diagram of $T^*$ can be explained by the fact that the coupling strength
between charge carriers, e.g.\ electrons, and the crystal lattice depends on the
velocity of the charge carriers. The faster the particles are, the smaller the
effect of deformation and the weaker the effective coupling potential.  This is
also an apparent explanation for the disappearance of superconductivity above
certain doping levels.  Namely, we propose that the PG phase is caused by
BCS-like pairing, but with pairs with momenta $\frac{|\vec k-\vec k'|}{|\vec
k_{\rm F}|} \ll 1$ that are close to each other. These pairs however do not
necessarily allow for
macroscopic coherence, i.e.\ long range order.

The appearance of pairings with finite center-of-mass momentum was suggested in
the sixties by Fulde-Ferell \cite{FF} and by Larkin-Ovchinnikov \cite{LO1,LO2}
independently and is nowadays referred to as FFLO phases. The pairs we study
here are of a different nature since their total momentum varies along the Fermi
surface. However, the form of these pairs naturally implies the existence of a
pair density wave (PDW), even though the pairing mechanism we propose here is
clearly different than the one usually discussed in the literature, see e.g.
~\cite{Agterberg,TL}. 

 The paper is organized as follows: We begin by discussing the
electron-phonon coupling in $\rm CuO_2$ and the resulting effective
electron-electron interaction in Section~\ref{sec:elph-int}.  Section \ref{sec:bcs} is dedicated to the
BCS gap equation arising from the presence of equal momenta electron pairs.
In Section \ref{sec:model1} we calculate distortion effects on the effective
electron-electron interaction in a tight-binding model. Next  we describe in 
Section~\ref{sec:res} numerical results showing that such Jahn-Teller type distortion
can give rise to non-zero electron-phonon coupling between pairs of electrons
with equal momenta and opposite spin and we discuss the resulting gap function and
pair wave densities.  
  In Section~\ref{sec:lingap} we study general pairings
  \((\vec k, \vec k')\) in an extended model with vanishing
  momentum transfer using the linearized gap equation.
  We show that close to the critical temperature exactly two distinguished
  pairings emerge, namely the \((\vec k, \vec k)\) pairing and the conventional
  \((\vec k, -\vec k)\) pairing, both with identical critical temperature~\(T^*\)
  satisfying the linear relation \eqref{eq:introT*}.
  However, the approximation of vanishing momentum transfer can only be justified
  for the \((\vec k, \vec k)\) case, as seen numerically in Section~\ref{sec:numext}. 
  There, we demonstrate the stability of $(\vec k,\vec k)$ pairs under certain conditions for non-vanishing
momentum transfers, also using the linearized gap equation. In particular, this
gives an example where $(\vec k,\vec k)$ is indeed the dominant pairing
mechanism. Furthermore the results for the pair wave function show explicitly
that the approximation of vanishing momentum transfer can only be justified for
\((\vec k, \vec k)\), but not for~\((\vec k, -\vec k)\).
The well-known derivations for the electron-phonon and Wegner effective
electron-electron interactions are briefly outlined in Appendix~A.

%%%%%%%%%%%%%%%%%%%%%%%%%%%%%%%%%%%%%%%%%%%%%%%%%%%%%%%%%%%%%%%%%%%%%%%%%%%%%%%%%%
%%%%%%
\section{effective electron-electron Interaction in ${\rm\bold{ CuO_2}}$}\label{sec:elph-int}

\begin{figure}[H]
  \begin{center}
    \includegraphics{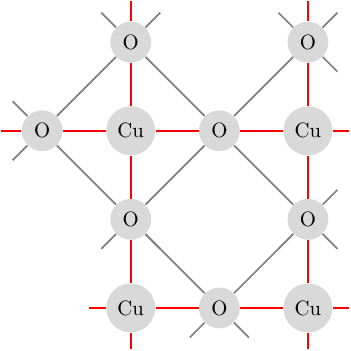}
 \end{center}
\caption{\normalsize Two dimensional $\rm CuO_2$ cubic lattice }\label{fig:lattice}
 \end{figure}

 As an example of a system which allows for the above described pairing mechanism, we consider a
 planar $\text{Cu}\text{O}_2$ lattice with volume $\Omega$ and square primitive cells composed of one copper atom and two oxygen atoms per unit cell, see Figure \ref{fig:lattice}. 
We are mainly interested in the interaction between Bloch electrons and lattice  phonons. 
Starting with the standard many-body Hamiltonian, the renormalization flow
  of Wegner \cite{W1, W2} yields an effective electron model where the electron-phonon interaction is replaced by an effective electron-electron
  interaction mediated by the phonons.  The
  leading-order effective Hamiltonian has the general form
  \begin{align}
 \Hel &= \sum_{\vec{k},n,\sigma} \eps_n(\vec{k}) c_{n\vec{k}\sigma}^\dagger c_{n\vec{k}\sigma} \nonumber
  \\&
+\sum_{\substack{\vec{k}n\sigma,\vec{k'}m\sigma'\\
  \vec{q}n'm'\vec{GG'}}} V^{nn'mm'}_{\sigma\sigma'}(\vec{k},\vec{k'},\vec{G},\vec{G'},\vec{q}) \;\nonumber
 \\&\qquad\qquad \cdot c_{n'\vec{k}+\vec{q}+\vec{G}\sigma}^\dagger c_{m'\vec{k'}-\vec{q}+\vec{G'}\sigma'}^\dagger c_{m\vec{k'}\sigma'} c_{n\vec{k}\sigma},
   \label{eq:effHam1}
\end{align}
 where $\eps(\vec{k})$ is the electronic dispersion relation, $\sigma,
\sigma'$ the electronic spins and $V$ the effective attractive interaction
between electrons with momenta $\vec k,\vec k'$ through a phonon with momentum
$\vec q$ and Umklapp vectors \(\vec G, \vec G'\).  It is worth noting that the
Wegner flow method has been previously used to study electron-phonon
interactions in other models, see e.g.\ \cite{ACGL}. 
\\ 
We are interested in possible pairing mechanism of electrons with momenta
$\vec{k}, \vec{k'}$, with $\frac{\abs{\vec k-\vec k'}}{\abs{\vec k_F}} \ll 1$
and both momenta are close to the Fermi surface with an effective interaction
mediated by phonons with low momenta~$\vec q$. However, in order to obtain an
explicitly solvable model, we further simplify this model by concentrating on
pairs with equal momenta. Thereby \eqref{eq:effHam1} can be restricted
to \(\vec k = \vec k'\), and  $\vec{k}+\vec{q}+\vec{G} =
\vec{k'}-\vec{q}+\vec{G'}$. Solving for the phonon momentum gives  $\vec q =
\frac{\vec G'-\vec G}2$. In the first Brillouin zone (FBZ) this has the trivial
solution~\(\bf q = 0\) and four further distinct solutions on the boundary,
\(\vec q = (\pi,0)\), \((0,\pi)\), and \( (\pi,\pm\pi)\), where 
the lattice constant~\(a = 1\) in natural units. 
Since we are interested in small momentum transfers, we focus on the case of
\(\bf q = 0\). It cannot be overemphasized that this should be considered as 
an approximation that captures the essential physical mechanism, whereas in an actual physical system any sufficiently small
 momentum transfer~\(\bf q\), and likewise any Bloch momentum pairs $\vec{k}, \vec{k'}$ that are sufficiently close to each
other on the scale of the Fermi momentum~$\vec{k}_F$, i.e.  $\frac{\abs{\vec
    k-\vec k'}}{\abs{\vec k_F}} \ll 1$, can contribute.     
In Sections~\ref{sec:lingap} and \ref{sec:numext}, we study more general pairings and potentials  by means of the linearized gap equations. There we provide some arguments and numerical evidence confirming the validity  of the approximations \(\vec k = \vec k'\) and \(\vec q = 0\)
in a simplified exemplary model.
   
Neglecting electron-electron Coulomb interactions, we obtain a reduced effective Hamiltonian of the form  
 \begin{align}\label{eq:eff-Ham-kk}
  \Heff &=\sum_{\vec{k}\sigma} \eps(\vec{k}) c_{\vec{k}\sigma}^\dagger c_{\vec{k}\sigma} +
\sum_{\substack{\vec{k},\sigma,\sigma'}} V(\vec{k})\;
  c_{\vec{k}\sigma}^\dagger c_{\vec{k}\sigma'}^\dagger c_{\vec{k}\sigma'} c_{\vec{k}\sigma}. 
\end{align}

 In the following sections, we explore the possible pair formation within this
 toy model. In the Appendix we briefly outline the standard derivation of the
 effective electron-electron interaction~\eqref{eq:effHam1} in the {\it rigid-ion} approximation.
 There one obtains for \eqref{eq:eff-Ham-kk} that
 \begin{align}\label{eq:effpotn}
   V(\vec k)&=-\sum_{\lambda} \frac 1 {\omega_\lambda(\vec 0)}  \abs{
   D_{\lambda}(\vec k)}^2.
 \end{align}
   where $\omega_\lambda(\vec 0)$ is the optical phonon energy at zero
   momentum, and the electron-phonon coupling~$D_{\lambda}(\vec k)$ is given by
  \begin{align} \label{eq:Dintro}
    D_{\lambda}(\vec k)=
    \Ii \sqrt{\frac{\hbar N_{\text{cell}}^{3}}{2 \omega_\lambda(0) \Omega^4}}
    \sum_{\tau}\sum_{\tilde {\vec G} \in \RL}&
    \vec e_{\lambda,\tau}(\vec 0) \SpDot \tilde {\vec G}
    \;
    \frac{ \hat v_{\text{ei}}^\tau(\tilde {\vec G})} {\sqrt{M_\tau}}\nonumber
     \\&\cdot\int_{\text{cell}}  \DInt[2] r\,\Ee^{\Ii \tilde {\vec G} \SpDot \vec r}
     \abs{u_{\vec k}(\vec r)}^2, 
   \end{align} 
  
  Here $N_{\text{cell}}$ is the number of primitive cells in a lattice of volume $\Omega$, $\tau$ runs over the atomic basis,  $M_\tau$ the mass of the $\tau$ ion and  $\hat v_{\text{ei}}^\tau$ the Fourier transform of  the {\it spin-independent} electron-ion potential, defined as
\begin{equation}
\hat{v}^\tau_{\rm ei}({\bf Q})= \int_\Omega \DInt[2]r \,v^\tau_{\rm ei}({\bf r})\Ee^{-\Ii{\bf
Q}\SpDot{\bf r}}
\end{equation}
 Moreover, $e_{\lambda,\tau}$ are the polarization vectors, while $u_k$ are the lattice periodic electronic wave functions and the integral is over the volume of the unit cell.
 
 It is worth noting that one obtains a similar expression for the effective
 electron-electron interaction between pairs $(\vec k,- \vec k)$ when  $\vec q=0$.   See the Appendix for details.

    %%%%%%%%%%%%%%%%%%%%%%%%%%%%%%%%%%%%%%%%%%%%%%%%%%%%%%%%%%%%%%%%%%%%
    
   \section{BCS approach to equal momentum pairing}\label{sec:bcs}
 
  %%%%%%%%%%%%%%%%%%%%%%%%%%%%%%%%%%%%%%%%%%%%%%%%%%%%%%%%%%%%%%%%%%%%%%
  We would like to emphasize that, as pointed out in \cite{HL}, any pairing, \(\vec k, \vec k'\) with $\eps(\vec k)=\eps(\vec k')$, can lead to the instability of the Fermi sea. Choosing equal momentum pairing allows us to obtain a gap equation that depends on only one momentum.
Let us now apply the usual BCS mean-field
 approach to~\eqref{eq:eff-Ham-kk}, with the gap function for equal momentum pairing  defined by
      \begin{equation}\label{eq:Delta}
    \Delta(\vec{k}) =    V(\vec{k}) \langle c_{\vec{k}\downarrow} c_{\vec{k}\uparrow} \rangle.
  \end{equation} 
Following standard arguments we obtain the gap equation 
\begin{equation}\label{eq:gap1}
    \Big( \frac{E_{\Delta}(\vec{k})}{\tanh\big(\frac{E_{\Delta}(\vec{k})}{2T}\big)}+ \frac{V(\vec{k})}{2}\Big)\Delta(\vec{k}) =  0,
  \end{equation}
  with
  \begin{equation}\label{eq:energy}
    E_{\Delta}(\vec{k}) = \sqrt{(\eps(\vec{k}) - \eps(\vec{k}_{\rm F}))^2 + |\Delta(\vec{k})|^2}.
  \end{equation}
The corresponding equation for the critical temperature~$T^*$,
\begin{equation}\label{eq:Tc1}
  \frac{E_{0}(\vec{k}_{\rm F})}{\tanh\big(\frac{E_{0}(\vec{k}_{\rm
  F})}{2T^*}\big)}= -\frac{V(\vec{k}_{\rm F})}{2},
  \end{equation}
reduces to the particularly simple relation 
  \begin{equation}\label{eq:T*}
    T^*=-\frac{V(\vec{k}_{\rm F})}{4}.
  \end{equation}
 The linear dependence on the coupling distinguishes this type of pairing from
 conventional superconductors. Here, the critical pairing temperature $T^*$
 is directly determined by the strength of the
 lattice deformation.  A particular weakness of our approach is the loss of phase dependence, since
 the solution of \eqref{eq:gap1} is determined only up to an arbitrary phase
 function $e^{i \theta(\vec{k})}$. 
  
It should be mentioned that in recent years the mathematical properties of conventional BCS theory have been intensively studied \cite{HHSS, FHNS, FHSS, HS1, HS2,FHaiLa,DeHaiSch} with sometimes rather surprising insights \cite{FHSchS,HSey}.
 %%%%%%%%%%%%%%%%%%%%%%%%%%%%%%%%%%%%%%%%%%%%%

 %%%%%%%%%%%%%%%%%%%%%%%%%%%%%%%%%%%%%%%%%%%%%%%%%%%%%%%%%%%%%%%%%
 \section{A tight binding model with Jahn-teller type distortion } \label{sec:model1}

%%%%%%%%%%%%%%%%%%%%%%%%%%%%%%%%%%%%%%%%%%%%%%%%%%%%%%%%%%%%%%%%%%%%%

As an illustrative example, we now augment the $\El{Cu}\El{O}_2$-model
from section~\ref{sec:elph-int} by a  Jahn-Teller type distortion. In
particular, we will show how such distortions give rise to attractive ${\bf
kk}$ interactions sufficient for the occurrence of BCS states with such 
pairings.  Our example of a lattice distortion is again intended to be a
{simplification} of the possible dynamically induced and thus usually localized
distortions. Consequently  many choices below will also be made with simplicity
and transparency of the resulting model in mind.

 \begin{figure}[H]
    \begin{center}
      \includegraphics{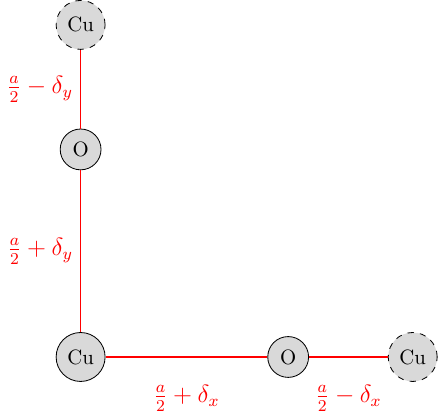}
\end{center}
\caption{\normalsize An example of a Jahn-Teller type distortion to $\El{Cu}\El{O}_2$. On the left,
  a unit cell with displacements $\delta_{x/y}$ from the positions of the
    oxygen atoms at the symmetry points \((a/2,0)\) and \((0,a/2)\) along the respective axes is shown. 
  }
\label{fig:latticedef}
  \end{figure}
\begin{figure}[H]
  \begin{center}
    \includegraphics{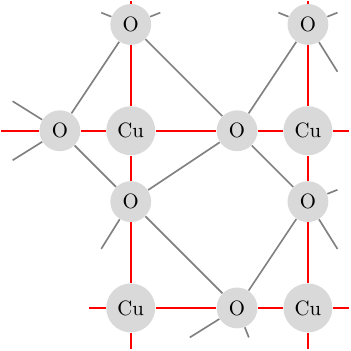}
 \end{center}

 \caption{The deformed lattice and bond structure resulting from the distortion
 of Figure~\ref{fig:latticedef}.} 
 \label{fig:latticedist}
 \end{figure}

We begin by statically distorting the two oxygens of each unit cell away from their
  symmetric equilibrium positions to \(r_{\El{O}^{(1)}}= (a/2+\delta_x, \, 0) \) and
  \(r_{\El{O}^{(2)}} = ( 0, \, a/2 + \delta_y)\).  Here we adopt dimensionless
  units with lattice spacing \(a=1\). The distortion length parameters
  \(\delta_x = \delta_y =: \delta\) are taken to be equal and 
  small compared to the lattice constant \(a\).  This geometry is shown in
  Figures~\ref{fig:latticedef}, \ref{fig:latticedist}.

 Next we calculate the electron-phonon coupling~\(D_\lambda(\vec k)\)
 using a tight-binding wave function 
 \begin{equation}
  \psi_{n,\vec k}(\vec r) = \frac{1}{\sqrt{N}} \sum_{j,\tau}  c_{\tau, \vec k}^n \Ee^{\Ii \vec k
  \SpDot  \vec R_{j}} w_{\tau}(\vec r - \vec R_{j \tau}),
\end{equation}
where \(N=3N_{\text{cell}}\) denotes the number of lattice ions. The
coefficients~\(c_{\tau, \vec k}^n\) are the \(n\)-th eigenvector of a hopping Hamiltonian
in the atomic basis
\begin{equation}
  {\mathcal H} = \begin{pmatrix} \varepsilon_{\rm Cu} & a_x & a_y \\
  a_x^* & \varepsilon_{{\rm O}_x} & c \\
  a_y^* & c^* & \varepsilon_{{\rm O}_y}  \\
\end{pmatrix} \label{eq:hophamilton}
\end{equation}
modeled after the lattice structure from Figure~\ref{fig:lattice},
with \(a_x := t_1 + t_1 \Ee^{- \Ii k_x}\), \(a_y := t_1 + t_1 \Ee^{- \Ii k_y}\)
and \(c := t_2 + t_2 \Ee^{\Ii k_x} + t_2 \Ee^{- \Ii k_y} + t_2 \Ee^{\Ii k_x -
\Ii k_y}\). The parameter \(t_1\) corresponds to horizontal and vertical \({\rm Cu}\)-\({\rm
O}\) hopping, while \(t_2\) is the amplitude for diagonal \({\rm O}\)-\({\rm O}\) hopping.

\begin{figure}[H]
   \begin{center}
   \includegraphics[width=210pt]{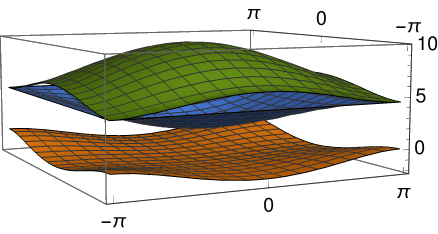}
   \\[2.0em]
   \includegraphics[width=190pt]{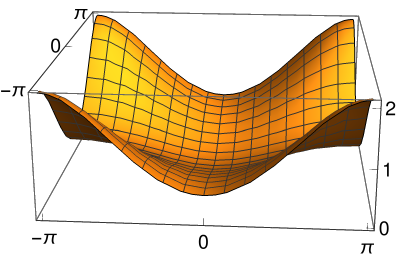}
   \end{center}
   \caption{\normalsize Dispersion relation in the tight-binding model
     with ~\(t_1=1.5~{\rm eV}\), \(t_2= 0.825~{\rm eV}\).  For the BCS model we
   consider only the lowest branch (bottom).}
   \label{fig:disptight}
 \end{figure}

 Typical values in $t_1$-units are 
 $\varepsilon_{\rm Cu}-\varepsilon_{\rm O}\approx 2.5 \text{ to }3.5 \, t_1$, 
 while $t_2\approx 0.5 \text{ to }0.6\,t_1$, with $t_1\approx 1.2 \text{ to }1.5
 \,{\rm eV}$
 \cite{PF, WMM}. Here we take $t_1=1.5 \,{\rm eV}, t_2=0.6 \, t_1$,
 $\varepsilon_{\rm Cu}=4.5\,{\rm eV}$, while  
 setting the oxygen ground state
energy \(\varepsilon_{\rm O}  \) to zero by a redefinition of the
Fermi energy. The resulting dispersion relation has three branches and is shown in
Figure~\ref{fig:disptight}.

 For the atomic wave functions we take Gaussians~\(w_\tau(\vec r) :=
{\mathcal N}_\rho \Ee^{-\vec r^2/(4\rho^2)}\), with width~\(\rho\) independent of the
atomic species~\(\tau\) and normalization~\({\mathcal N}_\rho^{-1} = \sqrt{2\pi\rho^2}\).
With this setup the resulting lattice-periodic wave functions are
\begin{equation}
  u_{n,\vec k}(\vec r) = {(2\pi)^2} \sum_{\tau} 
c_{\tau, \vec k}^n
\sum_{\vec G \in \RL}
  \Ee^{-\Ii \vec r \SpDot \vec G} \hat w(\vec k - \vec G) 
  \Ee^{-\Ii \vec R_\tau \SpDot (\vec k - \vec G)}
  \label{eq:u}
\end{equation}
where the integral from the Fourier representation~\(w(\vec r) := \int {\rm d}^2
q\; \Ee^{\Ii \vec r \SpDot \vec q} \hat w(\vec q) \) of the atomic wave
functions has already been carried out in combination with  
the lattice summation over~\(j\).
The reciprocal lattice (RL) sum can be performed numerically with an appropriate truncation 
or analytically using special functions.

Proceeding to the electron-phonon and
induced electron-electron interactions~\eqref{eq:Dintro}, we consider here only
the leading contributions from the smallest non-zero reciprocal lattice
components \( {\vec{\tilde G}} = (\pm 2\pi,0)\), \((0, \pm 2\pi)  \). 
For simplicity we will assume that electron-ion potential to be equal at these
momenta and independent of \(\tau\). Thus abbreviating \( v := \hat
v_{\text{ei}}^\tau(\pm 2\pi,0)\) we obtain
\begin{align} \label{eq:D}
    D_{\lambda}(\vec k)
    &\approx
    2 \Ii  v \sqrt{\frac{\hbar
    N_{\text{cell}}^{3}}{2 \omega_\lambda(\vec 0) \Omega^4}}
\left(
\sum_{\tau}
    \frac{ \vec e_{\lambda,\tau}(\vec 0) } {\sqrt{M_\tau}}
   \right)
   \SpDot 
   \begin{pmatrix}
     2\pi I_{\vec k}^a (2\pi ,0)
     \\
     2\pi I_{\vec k}^a (0 ,2\pi)
   \end{pmatrix}\nonumber
   \\&=
   \Ii v \sqrt{\frac{8 \pi^2 \hbar}{\omega_\lambda(\vec 0) \,
    \Omega }} \;
    \vec P_\lambda
   \SpDot 
   \begin{pmatrix}
     I_{\vec k}^a (2\pi ,0)
     \\
     I_{\vec k}^a (0 ,2\pi)
   \end{pmatrix},
   \end{align}
   cancelling \(\Omega =
   a^2 N_\text{cell}\) and recalling that we use natural units with~\(a = 1\).
   In the second equality we prepare for carrying out the mode sum over
     \(\lambda\) in the effective electron-electron
     interaction~\eqref{eq:effpotn} by introducing the polarization sum~\(\vec P_\lambda :=
   \sum_\tau M_{\tau}^{-1/2} {\vec e}_{\lambda, \tau}(\vec 0)\).
   Further note that the first equality we already replaced the reciprocal lattice sum over the
   electronic integral from \eqref{eq:Dintro} restricted to  \( {\vec{\tilde G}} =
     (\pm 2\pi,0)\), \((0, \pm 2\pi)  \) by twice the anti-symmetric part 
    \begin{align}\label{eq:Iamain}
    I_{\vec k}^a(\tilde{\vec G}) 
    = \Ii \int_{\text{cell}}  \DInt[2] r \, \sin(\tilde {\vec G} \SpDot \vec r) 
    \abs{u_{n,\vec k}(\vec r)}^2 
    \end{align}
     with
     \({\vec{\tilde G}} = (2\pi, 0)\), \((0,
       2\pi)\), as explained in more detail at the end of the appendix. 

   Guided by \eqref{eq:effpotn} we consider the two optical phonon modes with
   lowest energy. We denote their degenerate zero-momentum energy  by
   \(\omegalo := \omega_\lambda(\vec 0)\). 
   The above mode and atomic sum turns out to be independent of the choice of basis of
   the doubly degenerate polarization space. Further, using standard
   methods~\cite{Sol} to analyze the phononic structure of the present model,
   a basis of polarizations can be chosen with non-zero components purely in the \(x\)- or
   \(y\)-coordinate direction, respectively, yielding polarization sums \(\vec
   P_\lambda = (p,0)\) or \((0,p)\) for the respective phonon modes~\(\lambda\)
   for  some constant \(p \not = 0\).

  Note that an additional factor proportional to the volume~\(\Omega\) arises from our interpretation of \({\vec k}{\vec k}\) as an effective pairing. In particular, we consider $V(\vec k)$ as an approximation of the interaction between electrons with small relative momenta. Hence, the sums in \eqref{eq:effHam1} run over momenta in  a small neighborhood of $\vec k$. Overall this yields a factor proportional to the number of states in this neighborhood, which in turn is proportional to the volume $\Omega$.
   
   Any overall scale factors arising here are understood to be absorbed into the
   effective interaction constant \(v\).

 Altogether this yields a contribution to the effective  electron-electron
potential of 
\begin{equation}
   V(\vec k)   \approx 
   - \frac{8 \pi^2 \hbar | p v|^2 }
          {\omegalo^2}
    \left(
      \left| I_{\vec k}^a (2\pi ,0)\right|^2 +  
      \left| I_{\vec k}^a (0 ,2\pi) \right|^2
    \right).
    \label{eq:vfin}
\end{equation}
  
\section{Results for the Gap $\Delta$ and Pair-Wave Densities}
\label{sec:res}
\begin{figure}[H]
%  \vspace{0.9em}
  \begin{center}
   \includegraphics[width=210pt]{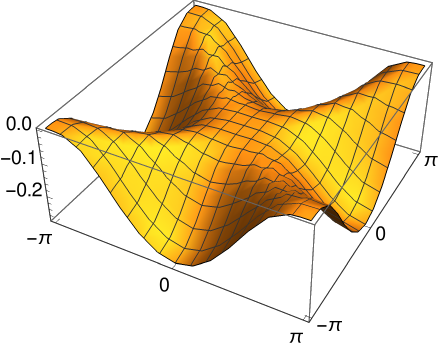}
 \end{center}
   \caption{ \normalsize The effective electron-electron potential~\(V\) for the lowest electron
     branch in the first Brillouin zone, shown
     in units of \(\hbar |p v|^2/\omegalo^2\) for distortion parameter 
   \(\delta = \rho = 0.05 a\).}
  \label{fig:V0}
\end{figure}

   We can now numerically demonstrate that the simplified
   distortion scheme from Figure~\ref{fig:latticedef} leads to a non-vanishing
   equal-momentum potential~\(V(\vec k)\). In Figure~\ref{fig:V0} the result for
   distortion parameter \(\delta = 0.05a\) is shown, with the remaining
   model parameters as in Section~\ref{sec:model1}. 
   The atomic wave function width~\(\rho = 0.05a\) is chosen rather small for
   simplicity, as for larger widths overlaps of neighboring atomic wave
   functions are longer negligible if we require that \eqref{eq:u} are well normalized.
   The numerics also confirm that \(V(\vec k)\) vanishes in the symmetric case with
   displacement~\(\delta=0\), whereas \(V < 0\) inside the first Brillouin zone
   if distortions are present.

\begin{figure}[H]
  \def\ww{190pt}
  \def\wl{-2.5em}
  \begin{center}
  \includegraphics[width=\ww]{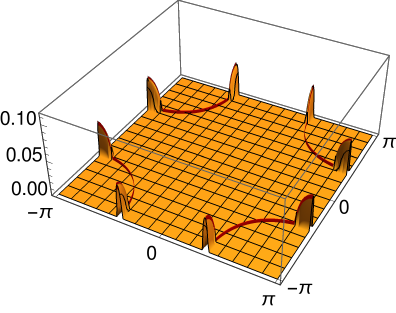}
\end{center}
\vspace*{\wl}
\hfill
 \(T \approx\ 0.9 T^* \)
  \\[0em]
  \begin{center}
    \includegraphics[width=\ww]{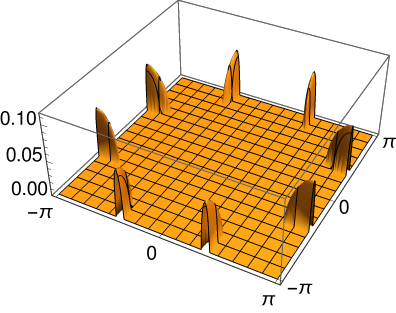}
\end{center}
\vspace*{\wl}
\hfill
     \(T \approx 0.75T^* \)
  \\[0em] 
  \begin{center}
   \includegraphics[width=\ww]{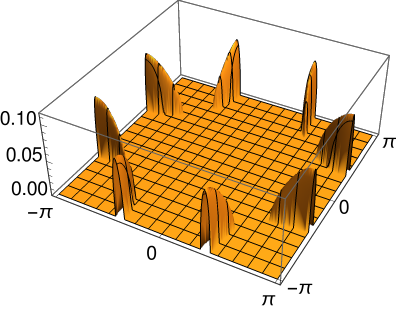}
\end{center}
\vspace*{\wl}
\hfill
   \(T \approx 0.5T^* \)
   \\[0em] 
  \begin{center}
   \includegraphics[width=\ww]{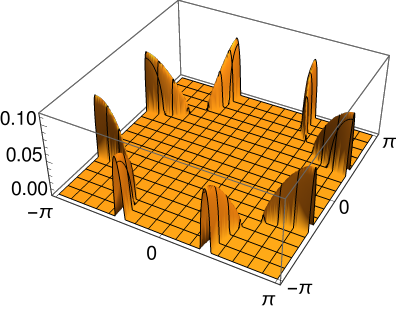}
\end{center}
\vspace*{\wl}
\hfill
   \( T \approx 0.4 T^* \)

   \caption{\normalsize Absolute value of the Gap function~\(\Delta\) in a tight-binding model
   with parameters~\(\rho = 0.05a\), \(\delta=0.05a\), \(t_1=1.5~{\rm eV}\),
 \(t_2= 0.8~{\rm eV}\). The Fermi surface with \(\mu = 0.8~{\rm eV}\) is indicated in red on the first gap plot. }
   \label{fig:gaptight}
 \end{figure}

Applying the results described in Section \ref{sec:bcs}, we obtain BCS states formed by \(\vec k
\vec k\)~pairs for temperatures~\(T\) below the critical temperature \(T^*\).
For this purpose we use the dispersion relation obtained as the lowest
eigenvalue of the hopping Hamiltonian~\eqref{eq:hophamilton}. 
The gap function~\(\Delta(\vec k)\) can then be obtained directly from
 \eqref{eq:gap1}. For \(T < T^*\) a non-vanishing gap starts to develop in the
vicinity of the maxima of \(V\) on the Fermi surface and extends to a
neighborhood of the full Fermi surface when lowering the temperature further as
shown in Figure~\ref{fig:gaptight}. 

\begin{figure}

  \begin{center}
 \includegraphics[width=240pt]{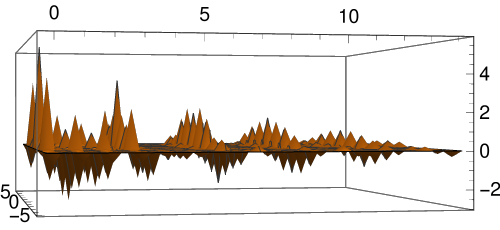}
  \\[0.5em]

  \includegraphics[width=240pt]{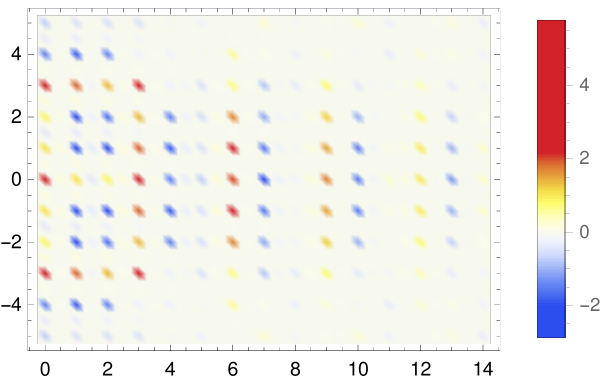}
\end{center}
 
  \caption{\normalsize Pair-wave density~\eqref{eq:pd} in configuration space for 
    order parameter \(\alpha(\vec k) = |\alpha(\vec k)|\) from
  Figure~\ref{fig:gaptight} at $T \approx\ 0.9 T^*$,
  calculated via Riemann sums with \(N=101\) support points in both coordinate
directions.}
   \label{fig:PDWs}
\end{figure}

\begin{figure}
 %\vspace{-0.43cm}
  \includegraphics[width=240pt]{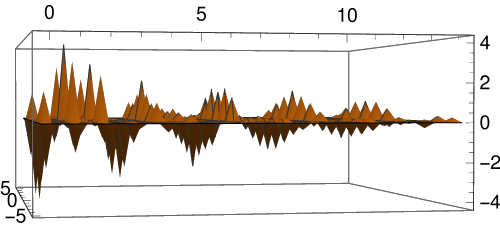}
  \\[0.5em]

  \includegraphics[width=240pt]{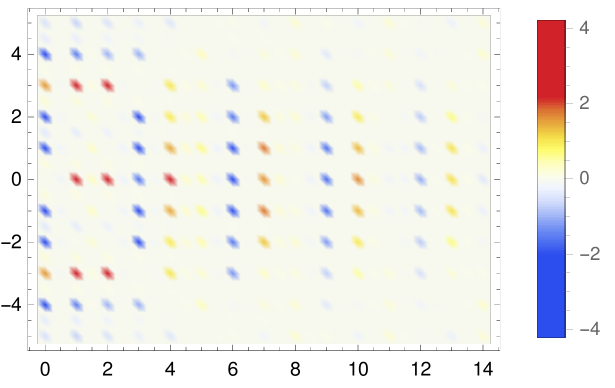}
 
  \caption{\normalsize Pair-wave density~\eqref{eq:pd} with additional d-wave-like
  phase~\(\alpha(\vec k) = \sigma(\vec k) |\alpha(\vec k)|\) at $T \approx\ 0.9
T^*$, \(\sigma(\vec k) := \sgn(k_x^2-k_y^2)\).}
\label{fig:PDWd}
\end{figure}

  It should be recalled here that \eqref{eq:gap1} yields only the absolute value
 \(|\Delta(\vec k)|\) of the gap function. On the other hand, the phase of the
 order parameter is not fixed by the present method, even to the extent that any
 choice of phase is consistent with this gap equation.
The pair density in position space evaluated
in the BCS state \(\Gamma\) from Section~\ref{sec:bcs}
is given by
\begin{equation}
\langle \psi_\uparrow(\vec r) \psi_\downarrow(\vec r)\rangle_\Gamma
  =
  \int_{\FBZ} \DInt[2]{k} \, \alpha(\vec{k}) \cos(2\vec{k}\cdot\vec{r})
  (u_\vec{k}(\vec{r}))^2  , 
\label{eq:pd}
\end{equation}
where by definition the Bloch field in the tight-binding model is \(\psi_\sigma(\vec r) = \int
\DInt[2]k \, \Ee^{\Ii \vec k \cdot \vec r} u_{\vec k}(\vec r) c_{\vec k \sigma}
\), we used the even parity symmetry of \(\alpha\) and \(u\) under \(\vec k
\leftrightarrow - \vec k\) and we note that \(u_{\vec k}\) is real-valued.
In Figures \ref{fig:PDWs} and \ref{fig:PDWd} we show some results for
two natural choices of the phase of the pairing order $\alpha(\vec{k})=\langle c_{\vec{k}\downarrow} c_{\vec{k}\uparrow} \rangle$.
In both cases, clear spatial
modulations of the pair density provide evidence for the emergence of pair
density waves (PDW) in the present model.

 Finally let us note that our model can easily be refined concerning various aspects. For 
 example, one could take into account the influence of the distortion on the
 hopping parameters~\(t_{1(2)}\) or include contributions of higher-order
 reciprocal lattice components \(\vec{\tilde G}\) in \eqref{eq:D}. Pursuing here would take us well beyond our
 present focus on the salient features of the proposed pairing mechanism. We
 hope that such questions will be explored in subsequent works.

 \section{Distinguished roles of \(\vec k,\vec k\) and \(\vec k,-\vec k\) among fully general pairings in the linearized gap equation} \label{sec:lingap}
 The linearized problem allows a direct comparison of the 
 standard Cooper pairing \((\vec k, -\vec k)\) with the fully generalized
 pairing \((\vec k, \vec k')\). 
 We begin by following the same steps and approximations as in Section~\ref{sec:model1}
 to obtain the electron-phonon potential for general pairs \((\vec k, \vec k')\)
 with momentum transfer \(\vec q = 0\)
 as
\begin{equation}
   V(\vec k, \vec k')   \approx 
   - \frac{8 \pi^2 \hbar | p v|^2 }
          {\omegalo^2}
          \begin{pmatrix}
       I_{\vec k}^a (2\pi ,0)\\
       I_{\vec k}^a (0 ,2\pi) 
     \end{pmatrix}
    \cdot
          \begin{pmatrix}
       I_{\vec k'}^a (2\pi ,0)\\
       I_{\vec k'}^a (0 ,2\pi) 
     \end{pmatrix},
    \label{eq:vfingen}
\end{equation}
with reduced effective Hamiltonian
 \begin{align}\label{eq:red-eff-Ham}
  \Heff &=\sum_{\vec{k}\sigma} \eps(\vec{k}) c_{\vec{k}\sigma}^\dagger c_{\vec{k}\sigma} +
   \!\!\!
   \sum_{\substack{\vec{k}\sigma \vec{k'}\sigma'}} V(\vec{k},\vec{k'})\;
  c_{\vec{k}\sigma}^\dagger c_{\vec{k'}\sigma'}^\dagger c_{\vec{k'}\sigma'} c_{\vec{k}\sigma}. 
\end{align}
We note that this is consistent with \eqref{eq:eff-Ham-kk}, where the latter is obtained by further reduction 
to quasi-free states supported on \((\vec k,\vec k)\) pairs only.
The potential has the general form
\begin{equation}
 - V(\vec k, \vec k') =  D_1(\vec k) D_1(\vec k') + D_2 (\vec k) D_2(\vec k'),
  \label{eq:potkkp}
\end{equation}
where \(D_2(k_y, k_x) = D_1(k_x, k_y) =: D(\vec k)\)
for the presently studied model.

 The linearized gap equation reads
 \begin{equation}
   \Delta = -\frac 1 2 L_\beta V \Delta
   \label{eq:lingap}
 \end{equation}
 with 2-body operator
 \begin{equation}\label{eq:L}
  L_\beta(\vec k, \vec k') = \frac{\tanh(\frac \beta 2 \epsilon_\mu(\vec k)  ) +
     \tanh(\frac \beta 2 \epsilon_\mu(\vec k'))}
     { \epsilon_\mu(\vec k) + \epsilon_\mu(\vec k') }
 \end{equation}
 and we abbreviate \(\epsilon_\mu(\vec k) := \epsilon(\vec k) - \mu\).
 Here we use the notation of \cite[Appendix~A]{HL}, where the reader can also
 find a succinct derivation and further explanations.

 For the toy model at hand, the product operator in~\eqref{eq:lingap} is a multiplication
operator and hence the eigenvalue problem becomes trivially solvable. The
critical~\(\beta^* = 1/T^*\) is defined by the emergence of a non-trivial
solution~\((\vec k, \vec k')\) of
\begin{equation}
  -\frac 1 2 L_{\beta^*}(\vec k, \vec k') V(\vec k, \vec k') = 1 
  \label{eq:tstar}
\end{equation}
and \(- \frac 1 2 L_\beta V < 1 \) for all \(\beta < \beta^*\).

\begin{figure}[H]
  \begin{center}
  \includegraphics[width=240pt]{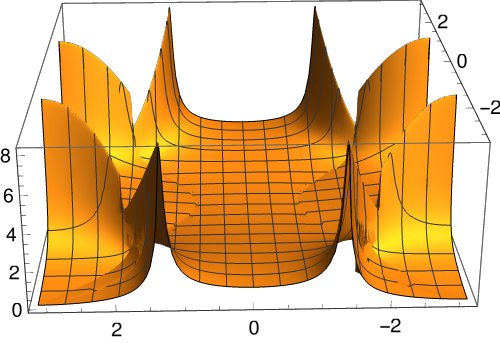}
\end{center}
  \caption{\normalsize The kernel maximum function \(M(\vec k) := \max_{k' \in \FBZ} 
      (- L_{\beta^*}(\vec k, \vec k') V(\vec k, \vec k'))\)  
      from the linear gap equation at
      \(\mu=0.85~\text{eV}\), where~\(\beta^*=100~\text{eV}^{-1}\).}
\label{fig:M}
\end{figure}

\begin{figure}[H]
\vspace{0.43cm}
\begin{center}
  \includegraphics[width=240pt]{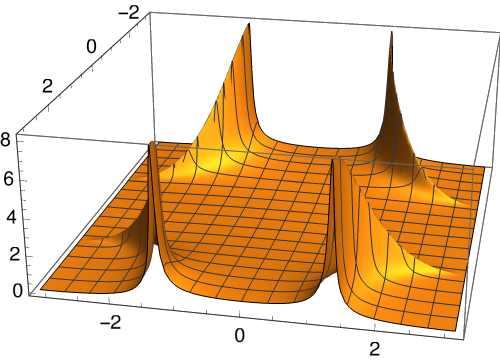}
\end{center}
  \caption{\normalsize The plot of the linear gap kernel \(-(LV)(\vec k_\text{max}, \vec k') \) as a
  function of \(\vec k'\) at one of the global maxima \(\vec k_\text{max}\) of
  \(M\) shows that exactly the two pairings \((\vec k_\text{max},\vec
  k_\text{max})\) and \((\vec k_\text{max},-\vec
  k_\text{max})\) emerge at the critical temperature. All other parameters are as in
Figure~\ref{fig:M}.}
\label{fig:LV}
\end{figure}

For simplicity, let us now adopt the perspective of fixing a
temperature \(T^*\) and then slowly turning on the potential (e.g.\ by 
a coupling constant).  From this perspective the global maxima of
the operator kernel from \eqref{eq:tstar} give the emerging dominant pairings.
For the parameters from Section~\ref{sec:res}, the numerics yield exactly the
conventional BCS pairings~\((\vec k,-\vec k)\) and the alternative pairings~\((\vec
  k, \vec k)\) studied in the present paper, as seen in Figures~\ref{fig:M},
  \ref{fig:LV}. 

One arrives at a similar conclusion by qualitative considerations: when the
kinetic kernel \(L_\beta\) provides the dominant scale, as for the present model
parameters, the first pairs to emerge are approximately located at the maximum
of the potential \(V\), when both momenta are on the Fermi surface
\(\epsilon_\mu(\vec k) = 0 =  \epsilon_\mu(\vec k')\) (see \cite{HL}).  In our
model, the maxima of the potential \(V(\vec k, \vec k')\) on the Fermi surface
are located at points exactly of the form \(\vec k' = \pm \vec k\). Thereby,
close to \(T^*\),  other types of pairing are excluded in our model.
This further motivates the study of the  \((\vec k, \vec k)\) pairing on the level 
of the fully non-linear gap equation in Section~\ref{sec:res} and confirms the necessity of considering alternative pairings. 

Finally, these two distinguished types of pairing can be compared analytically in the present model:
It is easily seen that \(V(\vec k, \vec k) = V(\vec k, -
\vec k)\) for all \(\vec k \in \FBZ\).
Similarly \(\epsilon_\mu(\vec k) = \epsilon_\mu(-\vec k)\) implies
\(L_\beta(\vec k, -\vec k) = L_\beta(\vec k, \vec k)\). Hence these two pairings correspond to exactly the same eigenvalue at the
level of the linearized gap equation \eqref{eq:lingap}. Thus they appear
also at exactly the same critical temperature. Numerically this can be
visualized by plotting \(M(\vec k) := \max_{k' \in \FBZ} 
(- L_{\beta^*}(\vec k, \vec k') V(\vec k, \vec k'))\), as shown in
Figure~\ref{fig:M}, and subsequently plotting \(-(L_{\beta^*}V)(\vec
k_\text{max} , \vec k')\) at one of the global maxima \(\vec k_\text{max}\) of
\(M\), as shown in Figure~\ref{fig:LV}. In the subsequent
Section~\ref{sec:numext} we will argue that this parity between \((\vec k, \vec
k)\) and \((\vec k, -\vec k)\) is not a true symmetry of nature.
Namely, we will demonstrate that the assumption \(\vec q = 0\) can be justified
for \((\vec k, \vec k)\) but fails for the conventional pairing, when also
interactions with non-vanishing momentum transfer are included.
    

\section{Emergence of equal momentum pairings for interactions with small momentum transfer.} 
\label{sec:numext}

Let us now consider the question of the stability of the observed
\((\vec k, \vec k)\)-pairings when interactions with
non-vanishing momentum transfers are included in the model. For this we return to the full Wegner interaction
\[
  H_{\text{int}} = 
\sum_{\substack{\vec{k}\sigma\vec{k'}\sigma'
  \vec{q}}} \hspace{-1em}
  V_{\sigma\sigma'}(\vec{k},\vec{k'},\vec{q}) \;\nonumber
 c_{n'\vec{k}+\vec{q}\sigma}^\dagger c_{m'\vec{k'}-\vec{q}\sigma'}^\dagger c_{m\vec{k'}\sigma'} c_{n\vec{k}\sigma},
\]
where Umklapp momenta are suppressed for notational simplicity.
As we are only interested in small \(\vec q\) and to remain comparable to our
main results, we will not amend our model to include a full phononic sector and
instead assume that the electron-phonon interaction is well approximated by
\(D_\lambda(\vec k, \vec q) \approx D_\lambda(\vec k)\) for small \(\vec q\) and
taken to vanish otherwise. To obtain a self-adjoint interaction we use an
appropriate extension of electron-phonon part from \eqref{eq:wegnerpot} to
nonzero \(\vec q\) 
given by 
\begin{align}
W(\vec k, \vec k', \vec q) 
 &:= \frac{1}{2}\sum_\lambda \left(
   D_{\lambda}(\vec k') D_{\lambda}(\vec k) 
 \right. \nonumber
   \\&\qquad \qquad \qquad
   \left.
   + D_{\lambda}(\vec k'-\vec q) D_{\lambda}(\vec k+\vec q)
 \right).
 \label{eq:potext}
\end{align}
Here we already used the approximation that \(\omega_\lambda(\vec q) \approx \omega_0 \not =
0\), constant and independent of the optical phonon mode \(\lambda\).
Hence the kinetic part from the Wegner interaction \eqref{eq:wegnerpot}
becomes independent of the phonon mode and the mode sum can be performed as
above.
On the other hand the matrix element of \(H_{\text{int}}\) providing the kernel
for the numerical study described below now has to be symmetrized under 
simultaneously exchanging \(\vec k \leftrightarrow
    \vec k'\) and \(\vec q \leftrightarrow - \vec q\) in order to conform to Fermi
    statistics, which yields
    \begin{align}
      V(\vec k, \vec k', \vec q) = - \frac {4 \omega_0 (\dd \dd' +
      \omega_0^2) }
      {(\dd^2 - \dd'^2)^2 + 4(\dd\dd'  + \omega_0^2)^2}  W(\vec k, \vec k', \vec q),
      \label{eq:sympot}
    \end{align}
    where \(\dd = \epsilon(\vec k + \vec q) -\epsilon(\vec k)\) and 
    \(\dd' = \epsilon(\vec k' - \vec q) - \epsilon(\vec k')\).
   % To anticipate the numerical results let us already note here that this
   % potential can in the above form be seen to contain singularities at \(\abs{\dd} \approx \abs{\dd'} \approx \omega_0\) which are either attractive or repulsive, depending on the manner in which they are approached.

\begin{figure}[H]
  %\null~\\
    \begin{center}
  \includegraphics[width=240pt]{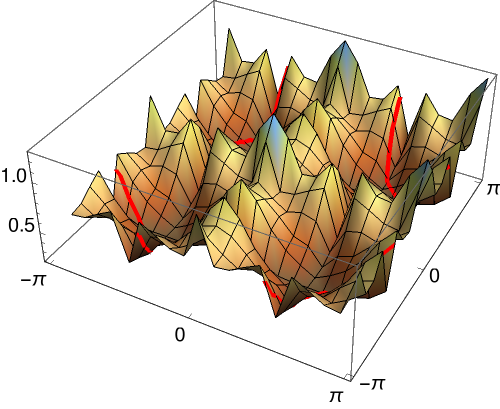}
\end{center}
  \caption{\normalsize Largest eigenvalue of \(-VL_\beta\) for \(\omega_0=0.5\;\text{eV}\) as function of~\(\vec K\) (other parameters
    as described in the text). Here and in the following figures we will indicate
    the Fermi surface for \(\mu = 0.85\;\text{eV}\) in red. The boundary points
  of the discretization will always only be included on the positive sides of
the corresponding axes. The plot meshes  are from now on matched to the discretization.}
  \label{fig:kmk}
\end{figure}

\begin{figure}[H]
  \begin{center}
  \includegraphics[width=240pt]{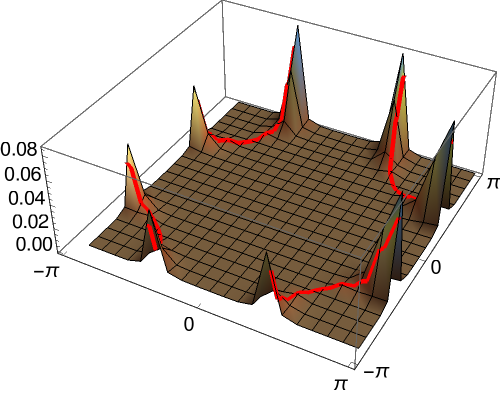}
\end{center}
  \caption{\normalsize Absolute square of the wave function for \(\vec K = 0\) in 
  Figure~\ref{fig:kmk} as function of \(\vec p\).
}
  \label{fig:kmkwf}
\end{figure}

\begin{figure}[H]
  \begin{center}
  \includegraphics[width=240pt]{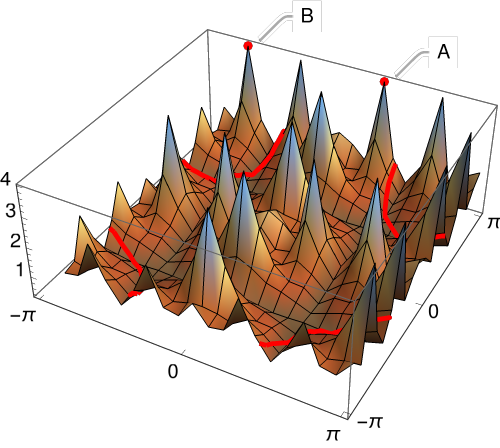}
\end{center}
  \caption{\normalsize
Largest eigenvalue for \(\omega_0=0.33\;\text{eV}\) as function of~\(\vec K\)
(other parameters as described in the text). The eigenvalues at ``A'', ``B'' and at
other similar peaks are
dominating over the eigenvalue at  the origin~\(\vec K = 0\). }
  \label{fig:kk}
\end{figure}

\begin{figure}[H]
  \begin{center}
  \includegraphics[width=240pt]{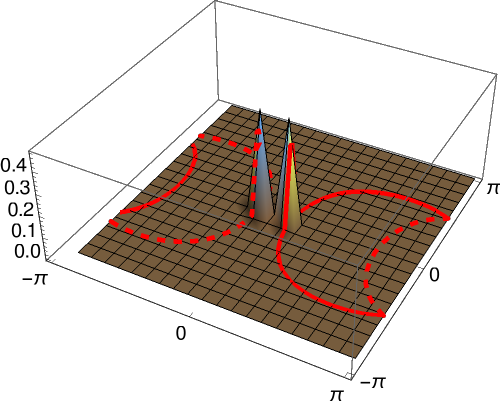}
\end{center}
  \caption{\normalsize
    Absolute square of the wave function  for \(\vec K=(\frac 3 {10}\pi, \pi)\) (point ``A`` in
    Figure~\ref{fig:kk}) as function of \(\vec p\). 
    Solid and dashed red lines show the Fermi surface for the two electron
    momenta \(\vec K + \vec
    p\) and \(\vec K - \vec p\), respectively.  The energy difference between the two
  peaks is proportional to \(\omega_0\).}
  \label{fig:kkwfa}
\end{figure}

\begin{figure}[H]
  \begin{center}
  \includegraphics[width=240pt]{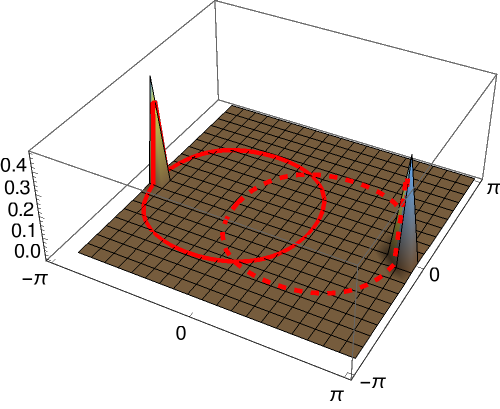}
\end{center}
  \caption{\normalsize
     Absolute square of the wave function  for \(\vec K=(-\frac 7 {10}\pi, \pi)\) (point ``B`` in Figure~\ref{fig:kk}) as function of \(\vec p\), showing that the
  wave function is concentrated near \((\pm \pi,0)\).}
  \label{fig:kkwfb}
\end{figure}

    We now study the spectrum of the operator \(-\frac 1 2 VL_\beta\) from the linearized gap equation \eqref{eq:lingap} using a suitable discretization. 
    As the linearized approximation of the gap equation is usually expected to
    be valid close to \(T^*\),
    the results from the main part of our paper suggest that the \(\vec k \vec k\)-pairing instability in the present model should appear close to the
    boundary of the first Brillouin zone. For this reason we use a
    discretization with periodic
    boundary conditions. To not accidentally suppress either the \((\vec k,
    -\vec k)\) or the expected novel \((\vec k, \vec k)\)-pairings,  we further
    carefully choose the discretization lattice to include both the origin and the boundary
    points of the form \((k_x, \pi)\)
    and \((\pi,k_y)\). For the numerical implementation we observe that at the
    level of the linearized gap equation \eqref{eq:lingap}, the various PDW-type pairing
    orbits~\((\vec k, \vec k') = (\vec K + \vec p, \vec K - \vec p)\) decouple.
    As in Section~\ref{sec:lingap} we identify the dominant pairing mechanism 
    from the largest eigenvalue of \(- \frac 1 2 V L_\beta\), which we 
    calculate here as function of \(\vec K\) together with the corresponding eigenfunctions. 
    For suitable parameters the numerical results shown in
    Figures~\ref{fig:kmk}--\ref{fig:kkwfb} provide further supporting evidence
    for our model.

   Due to the discretization approach the accessible lattice spacings are unfortunately limited by available computational resources.
    For the present calculation we choose a practical lattice
    discretizations of the first Brillouin zone with \(N_{\text{pt}}=20\) points
    per coordinate axis. We extend the potential via \eqref{eq:potext} to a
    \(\vec q\)-radius of \(2\) lattice spacings.
    The lattice spacing limits the ranges of numerically accessible
    temperatures \(T = \beta^{-1}\) and \(\omega_0\) from below, as the
    essential features of both the two-body operator~\(L_\beta\) and the
    Wegner potential have to be resolved with sufficient accuracy. 
    Both become less smooth as the corresponding parameter values are lowered.
    Due to these numerical limitations we choose here \(\beta = 50
      \;\text{eV}^{-1}\) and we lowered \(\omega_0\) very carefully starting
      from a physically very large value \(\omega_0 = 1\;\text{eV}\). Other model parameters are chosen as in
    Section~\ref{sec:model1}.
    Slowly lowering the phonon dispersion constant, we see that at larger
    \(\omega_0\) that the largest eigenvalues are at \(\vec K = 0\), corresponding to conventional 
    \((\vec k, -\vec k)\)-pairing, see
    Figure~\ref{fig:kmk}. The corresponding wave function as function of \(\vec p\) has the usual
     structure and is spread out over a close vicinity of the Fermi
    surface as seen in Figure~\ref{fig:kmkwf}.


  When \(\omega_0\) is further decreased additional peaks start to
    form, in particular at the boundary of the first Brillouin zone, as seen in
    Figure~\ref{fig:kk} for \(\omega_0 = 0.33 \; \text{eV}\). Already at this value of $\omega_0$ they dominate over the eigenvalue at \(\vec K = 0\).
    An inspection of the corresponding eigenfunctions
    reveals for the eigenvalue peak labeled ``A'' in Figure~\ref{fig:kk} a
    strongly concentrated wave-function near \(\vec p = 0\). Hence this
    yields \(\vec k \vec k\)-pairings as studied in this paper and thereby
    provides evidence supporting the approximation of vanishing momentum
    transfer.

 The additional peaks from Figure~\ref{fig:kk}
    can be explained by periodic boundary conditions. As an example,
    the wave function for the eigenvalue peak ``B'' is shown in Figure~\ref{fig:kkwfb}. Here we can see a strong concentration close to vectors of the halved reciprocal
    lattice on the boundary of the first Brillouin zone. This eigenvector is
    however physically equivalent to the eigenvector from point ``A'', as can be
    seen by translating both \(\vec K\) and \(\vec p\) by \((\pi, 0)\) and using
    periodicity. All remaining peaks can be similarly explained in terms of 
    ordinary ``A''-type \(\vec k\vec k\)-peaks by invoking the periodic boundary
    conditions.

Let us note that the electron energy difference between  \(\vec K + \vec
  p_{1/2}\) at the two peaks \(\vec p_{1/2}\) in Figure~\ref{fig:kkwfa} is
  comparable to \(\omega_0\). Hence can expect for physically small choices
  of \(\omega_0\) that the wave function is very well approximated by replacing
  it with just a single delta peak, which then yields exactly the model studied in
  the main part of this paper. On the other hand the results from
  Figures~\ref{fig:kmk} and \ref{fig:kmkwf} show that the same approximation is
  {\em not}\/ justified for the ordinary \((\vec k, -\vec k)\) pairing.

 We conclude this appendix by giving an explanation to the distinct behaviors of
 the \((\vec k,\vec k)\) and \((\vec k,-\vec k)\) wave functions. Let us
 consider the kinetic term in the symmetrized form of the Wegner potential
 from \eqref{eq:sympot}.  Now we note that there are configurations of \(\vec
 k\), \( \vec k'\)  and \(\vec q\) such that the absolute value of the parameter
 \(\varepsilon := \dd\dd' + \omega_0^2 \) becomes small.  In the regime
 \(\varepsilon \to 0\) we find the emergence of a Dirac delta potential
 \begin{align}\label{eq:limit}
      V(\vec k, \vec k', \vec q) 
      \overset{\varepsilon \to 0} \longrightarrow \mp 2  \pi \omega_0 \delta(\dd^2 - \dd'^2)
      W
     %  = \mp  \pi \delta(|\dd| - |\dd'|),
    \end{align}
    and this interaction is  an attractive or repulsive if the sign of
    \(\varepsilon\) is positive or negative, respectively.
    As the scattering processes most frequently take place close to the
    Fermi surface, the energy differences \(d = \epsilon(\vec k + \vec q) -
      \epsilon(\vec k)\) and \(d' = \epsilon(\vec k' - \vec q) -
    \epsilon(\vec k')\) tend to be close to zero.  Hence the case \(\varepsilon \geq 0\) is
   favoured, yielding a preference of nature for the attractive delta.

   However, the mechanism \eqref{eq:limit} can only contribute to the attractive interaction for 
   \((\vec k, \vec k)\) pairs and not for the conventional \((\vec k,-\vec k)\)
   pairs,
   since in the latter case we have \(d = d'\) and then \(\varepsilon \geq
   \omega_0^2 > 0\) prevents the realization of the limit in \eqref{eq:limit}. 

\section*{Conclusion}

We investigate a novel BCS-type pairing mechanism in which electron-electron
attraction is mediated by the interaction of low-momentum optical phonons and
Jahn-Teller-type lattice distortions. To keep the model as simple as possible
and allow for explicit calculations, we focus on the pairing of electrons with
equal momenta and give numerical evidence to validate this approximation. To demonstrate how this novel pairing mechanism can lead to
instability of the Fermi sea, we consider a particular distortion of a planar
${\rm CuO}_2$ lattice and using a tight-binding approximation, we numerically
calculate the BCS gap function in this case.  In the resulting toy model the Fermi sea is unstable towards equal momentum pairing below a certain
critical temperature $T^*$.  Due to the simplicity of the approach, which also
omits Coulomb interactions of electrons as well as density-density interactions
and exchange energies, we expect $T^*$ to represent not the actual critical
temperature describing macroscopic coherence, but the existence of localized
pairings such as the pseudogap. It is interesting to note that this appears to be the first microscopic model 
in which the pair density displays the characteristic features of a pair density wave (PDW).

\section*{Acknowledgement} 

C.H.\ is thankful to Mario Laux for his preliminary work on the model. 
C.H.\ also thanks Reinhold Kleiner and Niels Schopohl for fruitful discussions.
The authors also gratefully acknowledge the Leibniz Supercomputing Centre for providing computing time on its Linux-Cluster.

\section*{Appendix A: Electron-Phonon Coupling in ${\rm\bold{ CuO_2}}$}
 \label{app:elph-int}

In order to get an expression for the electron-phonon potential,  we follow the standard method outlined in many textbooks, e.g. \cite{Han, Sol}. However, we take into account the effect of reciprocal lattice vectors and Umklapp processes since they play important part in our discussion of electron pairs with equal momenta.

  Let $\Omega$ be the  volume of a lattice with $N_{cell}$ primitive cell, $N_e$ electrons and
let ${\bf r}$ denotes the position of an electron.   Using this notation, the electron-ion potential in the rigid ion approximation  can be written as

\begin{equation}
V_{\text{el-ion}}=\sum_{l=1}^{N_\text{e}} \sum_{j=1}^{N_\text{cell}}\sum_{\tau}
v^\tau_\text{ei} ({\bf r}_l-{\bf R}_{\tau j}),
\end{equation}

where ${\bf R}_{\tau j}$ is the position of the ``$\tau$" atom in the ``jth"
primitive cell and $\tau$ runs over the atomic basis. Note that $V_{\text{el-ion}}$
is periodic in the lattice parameter. Our
main assumption is  that $v^\tau_\text{ei}$ is spin independent and has a Fourier representation such that 
\begin{equation}\label{potnfourier}
v^\tau_\text{ei}({\bf r})=\dfrac{1}{\Omega} \sum_{{\bf Q}}\hat{v}^\tau_\text{ei}({\bf Q}) \Ee^{\Ii{\bf Q \SpDot r}},
\end{equation}
Note, that this assumption is fulfilled for example if $v^\tau_\text{ei}$ is periodic in the size of the lattice and bounded. 

In second quantization notation, this potential can be written in terms of the creation (annihilation)  operator $c^{\dagger}_{n {\bf k}\sigma} (c_{n {\bf k}\sigma}) $  of the one-particle electronic states characterized by the Bloch eigenstate $\psi_{n {\bf k}\sigma}$, with band index $n$ , wave number $k$  and spin $\sigma$, as follows

\begin{align}\label{eq:potin2quan}
V_\text{el-ion} = \sum_{j,\tau}\sum_{\substack{n,m\\\sigma\\\bf{k',k}\in
\FBZ}}&\Big\{\int_\Omega \DInt[2]{r}\, \psi_{n{\bf k'}\sigma}^*({\bf r}) v^\tau_\text{ei} ({\bf r}-{\bf R}_{\tau j})\psi_{m {\bf k}\sigma}({\bf r})\Big\}\nonumber
\\&\cdot c^{\dagger}_{n{\bf k'}\sigma}c_{m{\bf k}\sigma}.
\end{align}

Taking into account the displacement of the ions from their equilibrium position, the ionic position can be written as
\begin{equation}
{\bf R}_{\tau j}={\bf R}^{0}_{\tau j}+{\bf u}({\bf R}^{0}_{\tau j}),
\end{equation}
where ${\bf R}^{0}_{\tau j}$ is the equilibrium position of the $\tau j$ ion, while ${\bf u}_{\tau j}$ its displacement. 

Now for small displacements, the potential can be expanded  to first order as

\begin{equation}
v^\tau_\text{ei} ({\bf r}-{\bf R}_{\tau j})=v^\tau_\text{ei} ({\bf r}-{\bf R}^{0}_{\tau j})- { \bf \nabla}_{ {\bf r}} v^\tau_\text{ei} ({\bf r}-{\bf R}^{0}_{\tau j})\SpDot{\bf u}({\bf R}^{0}_{\tau j})+O(u^2).
\end{equation}

Inserting this expansion in (\ref{eq:potin2quan}), the first term gives the ``static" electron-ion interaction while the second is the electron-phonon interaction. Expressing the displacement of ions in terms of the phonon creation and annihilation operators $a^{\dagger}_{\lambda}({\bf q})$, $a_{\lambda}({\bf q})$, where $\lambda$ is the branch index and ${\bf q}$ is the phonon momentum taking values in the first Brillouin zone (FBZ), 
 the  electron-phonon interaction takes the form
\begin{align}
V_\text{el-ph}=-\sum_{{\bf q}\in \FBZ}&\sum_{\substack{n,m\\\lambda,\sigma}}\sum_{{\bf k',k}} D^{nm}_{\lambda, \sigma}({\bf k',k},{\bf q}) \nonumber
\\& c^{\dagger}_{n{\bf k'}\sigma}c_{m{\bf k}\sigma}\big( a_{\lambda}({\bf q})+a^{\dagger}_{\lambda}(-{\bf q}) \big).
\end{align}
Where the  electron-phonon coupling is given by  
\begin{align}\label{Dgeneral}
D^{nm}_{\lambda,\sigma}({\bf k',k, q})=&\sum_{j,\tau}\sqrt{\dfrac{\hbar}{2M_\tau
N_\text{cell} \omega_\lambda({\bf q})}}{\bf e}_{\lambda,\tau}({\bf q})
\Ee^{\Ii{\bf q}\cdot{\bf R}^{0}_{\tau j}}\nonumber
\\&\Big\{\int_\Omega \DInt[2]{r} \, \psi_{n{\bf k'}\sigma}^*({\bf r}) 
 { \bf \nabla}_{ {\bf r}} v^\tau_\text{ei} ({\bf r}-{\bf R}^{0}_{\tau j})\psi_{m {\bf k}\sigma}({\bf r})\Big\}.
\end{align}
Where ${\bf e}_{\lambda,\tau}({\bf q})$ are the polarization vectors extracted from the eigenvector of the Dynamical matrix corresponding to eigenvalue $\omega_\lambda({\bf q})$.
Using that $\psi_{m {\bf k}\sigma}$ are Bloch functions and  summing over $j$,  a simple calculation shows that the electron-phonon potential can be expressed in the terms of vectors in the reciprocal lattice (RL) as

\begin{align}\label{el-ph-eff}
V_\text{el-ph}=-\sum_{\substack{n,m,\lambda\\\sigma}} &\sum_{{\bf q},{\bf k}\in \FBZ}\sum_{\substack{{\bf G}\in \RL\\ {\bf k+q+G}\in \FBZ}}D^{nm}_{\lambda, \sigma}({\bf k},{\bf G},{\bf q})\nonumber
\\& \cdot c^{\dagger}_{n{\bf k+q+G}\sigma}c_{m{\bf k}\sigma}\big( a_{\lambda}({\bf q})+a^{\dagger}_{\lambda}(-{\bf q}) \big),
\end{align}
where the coupling is now given by
\begin{align}\label{D-G}
D^{nm}_{\lambda, \sigma}({\bf k},{\bf G},{\bf q})=&\sum_{\tau}\sqrt{\dfrac{\hbar
N_\text{cell}}{2M_\tau  \omega_\lambda({\bf q})}}e^{-\Ii{\bf G}\cdot{\bf R}^{0}_{\tau}}{\bf e}_{\lambda,\tau}({\bf q}) \nonumber
\\& \cdot\Big\{\int_\Omega \DInt[2]{r} \, \psi_{n{\bf k+q+G}\sigma}^*({\bf r}){ \bf \nabla}_{ {\bf r}} v^\tau_{\rm ei} ({\bf r})\psi_{m {\bf k}\sigma}({\bf r})\Big\}. 
\end{align}

 Using the  Fourier representation of the  electron-ion potential
 \eqref{potnfourier} and introducing the lattice periodic functions $u_{m {\bf
 k}\sigma}$ defined through $\psi_{m {\bf k}\sigma}({\bf r})=
 \dfrac{1}{\sqrt{\Omega}} \Ee^{\Ii{\bf k\SpDot r}} u_{m {\bf k}\sigma}$,  the
 coupling now takes the form 

\begin{align}\label{D-G-v-u}
D^{nm}_{\lambda,\sigma}({\bf k},{\bf G},{\bf q})=\Ii&\sum_{\tau, {\bf
Q}}\dfrac{1}{\Omega^2}\sqrt{\dfrac{\hbar N_{cell}}{2M_\tau  \omega_\lambda({\bf
q})}}\Ee^{-\Ii{\bf G} \SpDot {\bf R}^{0}_{\tau}}\nonumber
\\&\cdot\big({\bf e}_{\lambda,\tau}({\bf q})\cdot{\bf Q}\big) \hat{v}^\tau_{ei}({\bf Q})\nonumber
\\&\hspace{-1cm}\cdot\Big\{\int_\Omega \DInt[2]{r} \, \Ee^{\Ii{\bf Q}\SpDot{\bf r}}\Ee^{-\Ii{(\bf
q+G})\cdot{\bf r}} u_{n{\bf k+q+G}\sigma}^*({\bf r})u_{m {\bf k}\sigma}({\bf r})\Big\}. 
\end{align}

Finally, since the functions $u_{m {\bf k}\sigma}$ are lattice periodic (with trivial spin dependence), the integral over the volume can be reduced to integrals over the primitive cells. Therefore,  
 \begin{align}\label{D-final}
D^{nm}_{\lambda,\sigma}({\bf k},{\bf G},{\bf
q})&=\Ii\dfrac{N_\text{cell}}{\Omega^2}\sum_{\substack{\tau\\{\bf \tilde{G}}\in
  \RL}}\sqrt{\dfrac{\hbar N_\text{cell}}{2M_\tau  \omega_\lambda({\bf
q})}}\Ee^{-\Ii{\bf G}\cdot{\bf R}^{0}_{\tau}}\nonumber
\\&\cdot \big({\bf e}_{\lambda,\tau}({\bf q})\cdot {(\bf q+G+\tilde{G}})\big) \hat{v}^\tau_{ei}({\bf q+G+\tilde{G}})\nonumber
\\&\cdot \Big\{\int_{\text{cell}} \DInt[2]{r}\, \Ee^{\Ii{\bf \tilde{G}}\cdot{\bf
r}}u_{n{\bf k+q+G}\sigma}^*({\bf r})u_{m {\bf k}\sigma}({\bf r})\Big\},
\end{align}
where the integral is now over the volume of the primitive cell.

 Using the lowest order approximation of the Wegner flow \cite{W1,W2}, one obtains the following effective electronic Hamiltonian 
\begin{align}
 \Hel &= \sum_{\vec{k},n,\sigma} \eps_n(\vec{k}) c_{n\vec{k}\sigma}^\dagger c_{n\vec{k}\sigma} \nonumber
%  \\&
  + \hspace{-1em}\sum_{\substack{\vec{k}n\sigma,\vec{k'}m\sigma'\\
  \vec{q}n'm'\vec{GG'}}} \hspace{-1em} V^{nn'mm'}_{\sigma\sigma'}(\vec{k},\vec{k'},\vec{G},\vec{G'},\vec{q}) \;\nonumber
 \\&\qquad\qquad\qquad \cdot c_{n'\vec{k}+\vec{q}+\vec{G}\sigma}^\dagger c_{m'\vec{k'}-\vec{q}+\vec{G'}\sigma'}^\dagger c_{m\vec{k'}\sigma'} c_{n\vec{k}\sigma},
\end{align}
where
\begin{align}\label{eq:wegnerpot}
% \allowdisplaybreaks
 \hspace{-0.3cm} V^{nn'mm'}_{\sigma\sigma'}(\vec{k},&\vec k',\vec{G},\vec{G'},\vec q)
=
\sum_{\lambda}
   D_{\lambda \sigma'}^{m m'}(\vec k',\vec G',-\vec q)
   D_{\lambda \sigma}^{n n'}(\vec k,\vec G,\vec q)\nonumber
   \\&\quad\cdot
 \frac{
      \beta_{\lambda n n'}(\vec k,\vec G,\vec q)
      - \alpha_{\lambda m m'}(\vec k',\vec G',-\vec q)
 }{ (\alpha_{\lambda m m'}(\vec k',\vec G',-\vec q))^2 + (\beta_{\lambda n n'}(\vec k,\vec G,\vec q))^2},
\end{align}
%and
\begin{align}
   \alpha_{\lambda m m'}(\vec k,\vec G,\vec q) &=
\eps_{m'}(\vec k+\vec q+\vec G) -\eps_m(\vec k) +\omega_\lambda(\vec q),\label{eq:alpha}
\\\beta _{\lambda m m'}(\vec k,\vec G,\vec q) &=\eps_{m'}(\vec k+\vec q+\vec G)
-\eps_m(\vec k) -\omega_\lambda(\vec q).\label{eq:beta}
 \end{align}
  
 Eliminating the trivial spin dependence and restricting to a single band and optical phonon modes, for which
  \(\omega_\lambda(0) \not = 0\), and defining $D_{\lambda}^n(\vec k):=D^{nn}_{\lambda,\sigma}({\vec k},\vec 0,\vec 0)$ the electron-phonon coupling \eqref{D-final} yields the simple form \eqref{eq:Dintro}, where we also dropped the band index for convenience.

  Furthermore, using \eqref{eq:Dintro} along with \eqref{eq:wegnerpot}, \eqref{eq:alpha}, \eqref{eq:beta}
  and setting $V(\vec k)=V^{n}_{\sigma\sigma'}(\vec{k},\vec k,\vec 0,\vec 0,\vec 0) $ one obtains the effective electron-electron interaction  \eqref{eq:effpotn}.

 %%%%%%%%%%%%%%%%%%%%%%%%%%%%%%%%%%%%%%%%%%%%%%%%%%%%%%%%%%%%%%%%%%%%%%%%%%%%%%%%%%%%%%%%
Now let's take a closer look at the electron-phonon coupling \eqref{eq:Dintro}. %
Considering only the summation over \(\tilde {\vec G} \in \RL\) and assuming that the electron-ion potential \(v_{\text{ei}}^\tau\) is real and
  reflection symmetric, which implies that its Fourier coefficients also satisfy
  \(\hat v_{\text{ei}}^\tau(\tilde {\vec G)} = \hat v_{\text{ei}}^\tau(- \tilde
      {\vec G)} \). Together with the scalar product \(\vec
      e_{\lambda,\tau}(\vec 0) \SpDot \tilde{\vec G}\), we see that the prefactor of
    the electronic integral in \eqref{eq:Dintro} is
    anti-symmetric in \(\tilde {\vec G}\). But this means that only the
    anti-symmetric parts of the electronic integrals
    \begin{align}\label{eq:Ia}
    I_{\vec k}^a(\tilde{\vec G}) 
    = \Ii \int_{\text{cell}}  \DInt[2] r \, \sin(\tilde {\vec G} \SpDot \vec r) 
    \abs{u_{\vec k}(\vec r)}^2 
    \end{align}
    can yield non-vanishing contributions to \(D_\lambda(\vec k)\).
    It is easy to see that in the case a ``perfect" $\rm CuO_2$ crystal, this integral vanishes. However, a Jahn-Teller type distortion, where the symmetry of the crystal is broken, can cause the integral \eqref{eq:Ia} to be  non-zero. Resulting in a non-zero electron-phonon coupling and the possible formation of equal momenta electron pairs.   
  %%%%%%%%%%%%%%%%%%%%%%%%%%%%%%%%%%%%%%%%%%%%%%%%%%%%%%%%%%%%%%

% Fakesection: backmatter

%%%%%%%%%%%%%%%%%%%%%%%%%%%%%%%%%%%%%%%%%%%%

\end{document}